\documentclass[journal=nalefd,manuscript=letter]{achemso}

\usepackage{color}
\usepackage{graphicx}
\usepackage{dcolumn}
\usepackage{bm}
\usepackage{braket}
\usepackage{dsfont}
\usepackage{color}

\title{Chemisorption of water on the surface of silicon microparticles measured by DNP-enhanced NMR}

\author{Mallory L. Guy}
\author{Kipp J. van Schooten} 
\author{Lihuang Zhu}
\author{Chandrasekhar Ramanathan}
\email{chandrasekhar.ramanathan@dartmouth.edu}

\affiliation{Department of Physics and Astronomy, Dartmouth College, Hanover, New Hampshire 03755, USA}


\begin{document}

\begin{abstract}
We use dynamic nuclear polarization (DNP) enhanced nuclear magnetic resonance (NMR) at liquid helium temperatures to directly detect hydrogen attached to the surface of silicon microparticles.   The proton NMR spectrum from a dry sample of polycrystalline silicon powder (1--5 $\mu$m) shows a distinctively narrow Lorentzian-shaped resonance with a width of 6.2 kHz, indicative of a very sparse distribution of protons attached to the silicon surface.  These protons are within a few atomic monolayers of the silicon surface.  The high sensitivity NMR  detection of surface protons from low surface area ($0.26-1.3$ m$^2$/g) particles is enabled by an overall signal enhancement of 4150 over the room temperature NMR signal at the same field.  When the particles were suspended in a solvent with 80\% H$_2$O and 20\% D$_2$O, the narrow peak was observed to grow in intensity over time, indicating growth of the sparse surface proton layer.  However, when the particles were suspended in a solvent with 20\% H$_2$O and 80\% D$_2$O, the narrow bound-proton peak was observed to shrink due to exchange between the surface protons and the deuterium in solution.  This decrease was accompanied by a concomitant growth in the intensity of the frozen solvent peak, as the relative proton concentration of the solvent increased.  When the particles were suspended in the organic solvent hexane, the proton NMR spectra remained unchanged over time.  These results are consistent with the known chemisorption of water on the silicon surface resulting in the formation of hydride and hydroxyl species. 
Low-temperature DNP NMR can thus be used as a non-destructive probe of surface corrosion for silicon in aqueous environments. This is important in the context of using silicon MEMS and bioMEMS devices in such environments,  for silicon micro- and nano-particle MRI imaging agents, and the use of nanosilicon for splitting water in fuel cells.
\end{abstract}

\section{Introduction} 
Water molecules are known to chemisorb onto the surface of silicon, resulting in the formation of both Si-H and Si-OH groups at the silicon surface \cite{Ciraci-1983,Schaefer-1985,Haneman-1987}.  It has been suggested that such oxidation of silicon in aqueous environments can lead to degradation of silicon-based bio-MEMS (micro-electro-mechanical-systems) devices \cite{Grayson-2004,Iliescu-2012} that have not been adequately encapsulated \cite{Ciacchi-2008}.   The interaction of water with the silicon surface is also important for the potential use of silicon nanoparticles as tracers in biomedical applications.  Silicon nano- and micro-particles have been shown to be promising biomedical magnetic resonance imaging agents \cite{Cassidy-2013b,Whiting-2015}.  The silicon-29 nuclear spins can be hyperpolarized using dynamic nuclear polarization (DNP) at cryogenic temperatures \cite{Dementyev-2008} and the long spin-lattice relaxation times of the silicon-29 spins in the the solid ensures that the polarization is preserved when the particles are warmed to room temperature, potentially offering significant advantages over dissolution DNP methods \cite{ArdenkjaerLarsen-2003}.  The silicon surface can also be bio-functionalized for targeted molecular imaging applications \cite{Aptekar-2009}.  In the presence of a potassium hydroxide (KOH) activator, silicon nanoparticles have also been used to split water for hydrogen generation \cite{Erogbogbo-2013}.

While nuclear magnetic resonance (NMR) spectroscopic techniques have long been used to provide both chemical and structural information at the atomic level, they have previously not been sensitive enough to detect most surface chemical species.  The relative insensitivity of NMR is due to the small Zeeman energies ($\sim$ $\mu$eV) available at typical laboratory magnetic fields.  Since the magnetic moment of the electron is 658 times larger than that of the proton (hydrogen nucleus), the equilibrium polarization of the electron spins is correspondingly larger.  During DNP, microwave irradiation of a coupled electron-nuclear system at or near the electron spin Larmor frequency results in a transfer of polarization from the electrons to the nuclear spins, significantly enhancing the nuclear spin polarization.   

While much higher electron (and thus nuclear) polarizations can be achieved at liquid helium temperatures, the robust magic angle sample spinning (MAS) needed for the use of standard high-resolution solid state NMR techniques with DNP has necessitated operation at liquid nitrogen temperatures \cite{Becerra-1993,Bajaj-2003}.  This technology, which has recently been commercialized \cite{Rosay-2010}, has rapidly expanded the application of DNP.  Such DNP-enhanced MAS NMR spectroscopy has emerged as an important methodological tool to study chemistry and structural ordering at surfaces and interfaces \cite{Lesage-2010,Zagdoun-2011,Lelli-2011,Rossini-2012,Rossini-2013,Kobayashi-2013,Sangodkar-2015,Piyeteau-2015,Perras-2016,Johnson-2016}.  In these experiments a free radical polarizing agent is explicitly added into the solvent, or impregnated into the surface of the particles.  Recently, Sangodkar {\em et al}.\ used DNP-enhanced MAS NMR at liquid nitrogen temperatures to study the competitive adsorption of organic molecules and water at the surface of of silicate particles \cite{Sangodkar-2015}.  

Surfaces have also been studied with static DNP NMR using electron spins that are endogenous to the material, such as the dangling bonds present at the silicon-silicon dioxide interface \cite{Dementyev-2008,Cassidy-2013a}  or those present at the surface of diamond nanoparticles \cite{Rej-2015,Bretschneider-2016}.  At low field, Rej {\em et al}.\ demonstrated the use of DNP to measure the presence of adsorbed species at the surface of diamond nanoparticles \cite{Rej-2016}.  Static DNP enhancements performed at liquid helium temperatures allow for significantly greater sensitivity, albeit potentially at the cost of spectral resolution.  It should be noted, however, that experiments in which the MAS rotor is kept at liquid nitrogen temperatures while the sample is further cooled to liquid helium temperatures have been demonstrated \cite{Thurber-2016}.

\begin{figure*}
\begin{center}
\includegraphics[width = 0.9\textwidth]{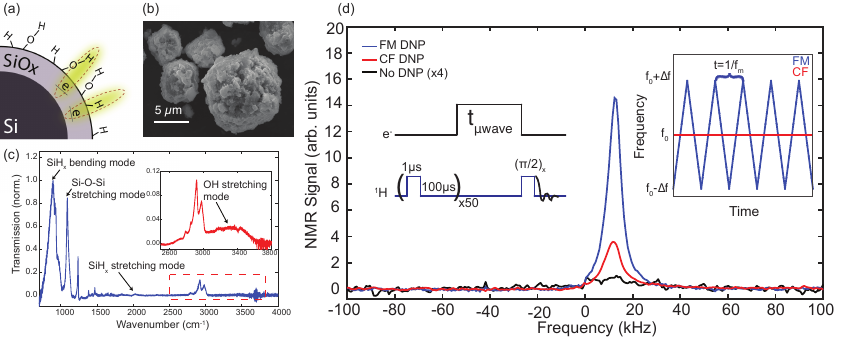}
\caption{(a) Schematic illustration of the dry particles with hydride and hydroxyl groups attached to the surface as well as the presence of silicon dangling bonds; (b) SEM micrograph of the particles; (c) FT-IR data from the dry particles indicating the presence of multiple Si-H modes, as well as an -OH mode; (d) Proton NMR spectra from the dry silicon powder obtained under different experimental conditions at 4.2 K: without DNP; with constant frequency (CF) DNP; and with frequency-modulated (FM) DNP.  A modulation amplitude ($\Delta f$) of 30 MHz and a modulation frequency ($f_\mathrm{m}$) of 10 kHz were used for the FM DNP spectrum.  The DNP enhancements measured are 
$\epsilon_\mathrm{CF}=14.1$ and $\epsilon_\mathrm{FM}=58.2$.  A build-up time of 256 s was used in all experiments.  The inset depicts the difference between the CF and FM DNP methods.}
\label{fig:Fig1}
\end{center}
\end{figure*}

In this paper we use static DNP-enhanced NMR at liquid helium temperatures to directly detect the protons attached to the surface of silicon microparticles.  For a dry silicon powder sample (1--5 $\mu$m polycrystalline silicon - Alfa Aesar) the DNP-enhanced proton NMR spectrum is a relatively narrow Lorentzian-shaped resonance with a width of 6.2~kHz, indicative of sparsely distributed protons that are isolated from each other.  These protons are within a few atomic monolayers of the silicon surface. The high sensitivity DNP NMR signal detection is enabled by an overall signal enhancement of 4150 over the room temperature NMR signal at the same field.   At cryogenic temperatures, molecular motions are frozen and the dominant spin-spin interaction is the magnetic dipolar coupling between the spins.  The linewidth of the measured NMR spectra indicate the strength of the magnetic dipolar couplings, which in turn are determined by the spatial distribution of the spins.   We also probe the interaction of the silicon powders with three different solvents by first measuring DNP-enhanced proton NMR spectra from the freshly suspended particles, and then repeating the experiments five days later to characterize any signal changes induced by solvent particle interactions.  The magnitude of the narrow Lorentzian peak was observed to change over time when the particles were suspended in aqueous solvents, but remained unchanged when the particles were suspended in hexane. 
  
\section{Results and Discussion} 

\subsection{DNP of dry particles}

Figure \ref{fig:Fig1}(a) shows a cartoon of the surface of the dry silicon particles with hydride and hydroxyl groups attached to the surface as well as the presence of silicon dangling bonds.  It is known that the surfaces of semiconductors such as silicon always have a significant interfacial density of dangling bond defects due to the lattice mismatch between the silicon and the silicon dioxide that grows on the surface cleavage planes. 
For silicon micro and nanoparticles, the unpaired electrons of the dangling bond are naturally localized at the surface of the particle, while the interior core of the particles remains relatively defect free \cite{Dementyev-2008}, which benefits the NMR study of surfaces with DNP.  Figure \ref{fig:Fig1}(b) shows an SEM micrograph of the silicon particles that typically range in diameter from 1--5 $\mu$m, showing their relatively heterogenous structure.  The estimated surface area for such particles is on the order of $0.26-1.3$ m$^2$/g (assuming spherical particles).  X-ray diffraction of these microparticles suggested that about 80\% of the sample is amorphous while 20\% is crystalline \cite{Dementyev-2008}.
Figure \ref{fig:Fig1}(c) shows the FT-IR spectrum measured from the dry silicon powder indicating the presence of multiple Si-H$_\mathrm{x}$ modes, as well as an -OH mode.  This suggests the presence of both hydride and hydroxyl groups at the silicon surface, potentially formed by exposure to atmospheric moisture. 

\begin{figure}
\includegraphics[width = 0.45\textwidth]{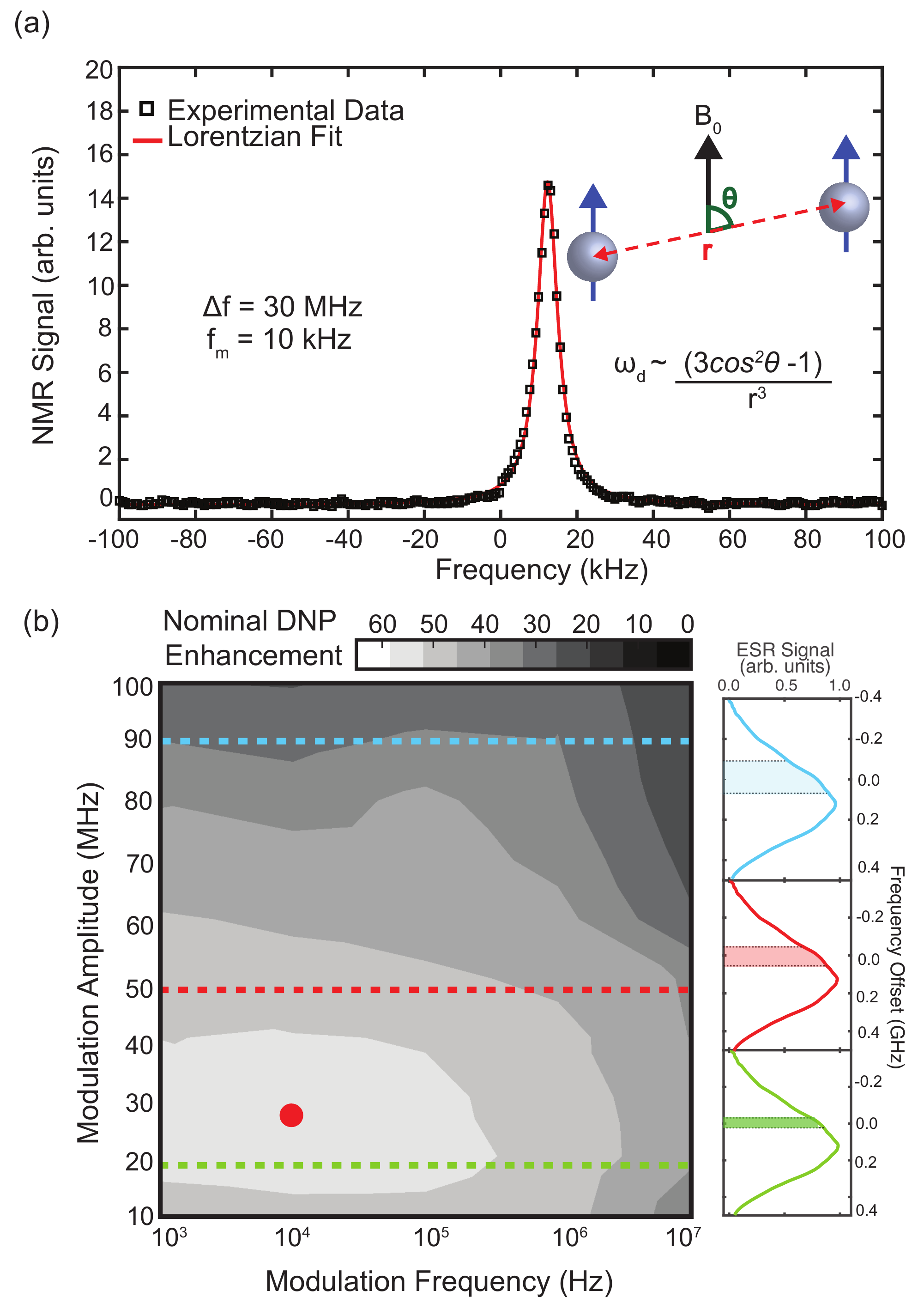}
\caption{(a) FM-DNP proton NMR spectrum (black squares) and the best Lorentzian spectral fit (red line - 6.2 kHz width).   The inset on the right shows a cartoon illustrating how the strength of the magnetic dipolar coupling between two nuclear spins depends on their geometry. (b) Interpolated surface plot shows the nominal proton DNP enhancement $\epsilon$ obtained for the dry powder as a function of modulation amplitude ($\Delta f$) and modulation frequency ($f_\mathrm{m}$). The red dot corresponds to the conditions under which the spectrum in (a) were acquired ($\Delta f = 30$ MHz and $f_\mathrm{m} = 10$ kHz). The plots on the right show the longitudinally-detected ESR spectrum of the silicon particles acquired at 94 GHz, as well as the regions of the ESR line swept while using three different modulation amplitudes.  The standard deviation of the signal-free region of the spectrum indicates the noise level in the experiment.}
\label{fig:Fig2}
\end{figure}

Figure \ref{fig:Fig1}(d) shows the proton NMR spectra obtained under different experimental conditions at 4.2 K: without DNP; with constant frequency (CF) DNP; and with frequency-modulated (FM) DNP.  The DNP enhancement is defined as $\epsilon = S_{\mathrm{\mu w \: on}}/S_{\mathrm{\mu w \: off}} - 1$, where $S$ is the signal measured in the NMR experiment (see Methods).  The spectra were obtained with a single acquisition and showed DNP enhancements of $\epsilon_\mathrm{CF}=14.1$ and $\epsilon_\mathrm{FM}=58.2$ (corresponding to an overall enhancement of $58.2 \times 71.4 = 4150$ over the 300 K room temperature signal).  
The NMR experiments were performed at 4.2 K in a 3.34 T super-widebore superconducting NMR magnet equipped with a Janis continuous-flow NMR cryostat, and a Bruker Avance AQX NMR spectrometer.  The electron spin Larmor frequency is close to 94 GHz and the proton Larmor frequency is 142 MHz.  The millimeter waves are generated using a previously described millimeter wave source \cite{Guy-2015} and the techniques used to produce the frequency modulation of the millimeter waves are described in the Methods section.  Before each measurement, a 50-pulse saturation train was applied to the spins (inset of Figure \ref{fig:Fig1}(d)) to destroy the equilibrium nuclear magnetization.  Following the saturation train, the nuclear spin polarization was allowed to build up --- either in the presence (DNP) or absence (no DNP) of millimeter wave irradiation --- for a duration of 256 s, and the NMR signal measured using a $\pi/2$ readout pulse.  The duration of the $\pi/2$ pulse is 1.5 $\mu$s, corresponding to an excitation bandwidth of $\sim 170$~kHz.

Figure \ref{fig:Fig2}(a) shows that the proton NMR spectrum following FM DNP is very well fit by a 6.2 kHz wide Lorentzian-shaped resonance line.  The width of a magnetic resonance spectral line is typically determined either by the inhomogeneity of the magnetic field across the sample or the strength of the magnetic dipolar couplings experienced by the spins \cite{Haeberlen}.  The strength of the magnetic dipolar coupling 
($\omega_\mathrm{d}$) between two spins is inversely proportional to the distance between them.  At high magnetic fields,  the coupling also has an angular dependence as shown in the inset of Figure \ref{fig:Fig2}(a).  
In proton-dense solids, typical dipolar-determined resonance linewidths are on the order of 50-100 kHz.  
The proton spectra measured from the silicon powder is similar to those observed in previous NMR studies of hydrogenated-amorphous silicon films in device-quality wafers  (with 8--20 at.\ \% hydrogen)\cite{Carlos-1980,Reimer-1980,Reimer-1981,Carlos-1982,Zumbulyadis-1986,Baum-1986,Gleason-1987,Wu-1996,Baugh-2000,Baugh-2001}.  These studies identified two dominant features of the spectrum that were present at temperatures ranging from room temperature down to liquid helium temperatures:  a relatively narrow line with a width of about 4 kHz associated with sparsely distributed silicon-monohydride species (with $\sim 4$ at.\ \% hydrogen \cite{Baum-1986}) and a broader component on the order of 25-35 kHz associated with clusters of the mono-hydride.  The spacing between protons contributing to the narrow spectral line was estimated to be 6-8 \AA  \cite{Wu-1996}.  Here, the observed NMR spectrum contains contributions from  both hydride and hydroxy protons, as the spectral features of both species are likely to overlap.

Figure \ref{fig:Fig1}(d) shows that frequency modulation of the microwaves significantly improve DNP efficiency.
FM DNP has previously been shown to improve signal enhancement by factors of 2--10, depending on the sample and the experimental conditions \cite{Cassidy-2013b,Hovav-2013,Bornet-2014,Guy-2015}, in  systems that hyperpolarize via the cross effect \cite{Wollan-1976,Hovav-2013}.  The cross effect is a 3-spin process involving one nuclear spin and two dipolar-coupled electron spins whose Larmor frequencies differ by the nuclear Larmor frequency.  Millimeter wave irradiation at either of the electron spin resonance frequencies leads to a polarization enhancement of the nuclear spins.  If the electron spin resonance line has a significant degree of inhomogeneous broadening, constant frequency millimeter wave irradiation only excites a relatively small subset of electron spins in the sample.  In this case modulation of the millimeter wave frequency increases the number of electron spins contributing to the DNP process, further increasing the NMR signal.

The interpolated surface plot of Figure \ref{fig:Fig2}(b) shows the nominal DNP enhancement ($\epsilon$) measured for the surface protons in the dry powder as a function of the modulation parameters ($\Delta f$ and $f_\mathrm{m}$) defined in Figure \ref{fig:Fig1}(d).  The low signal-to-noise ratio (SNR) for the thermal signal (no DNP) leads to a significant degree of uncertainty in the reported DNP enhancement -- see the Supplementary Information for the fit used to calculate the enhancement factors.  The circular red dot indicates the conditions under which the spectrum shown in Figure \ref{fig:Fig2}(a) was acquired.  The right side of Figure \ref{fig:Fig2}(b) shows a longitudinally-detected (LOD) ESR spectrum of the silicon particles acquired at 94 GHz using a similar scheme to that described previously \cite{Granwehr-2007}, as well as the regions of the ESR line swept using three different modulation amplitudes.   The powder sample used in the LOD ESR experiments was significantly larger than those used in the DNP experiments, potentially leading to a broader line. The DNP enhancement with frequency modulation is seen to be at least twice  as large ($\epsilon > 30$) as the enhancement measured at constant frequency ($\epsilon = 14.1$) over the entire range of modulation parameters used.  Maximal DNP enhancements were obtained for modulation amplitudes in the range from $\Delta f = $ 20--40 MHz and modulation frequencies $f_\mathrm{m} < 200$ kHz.  Vega and co-workers have suggested that the modulation amplitude dependence is dominated by inhomogeneities in the ESR line, while the modulation frequency dependence is determined by the electron spin spin-lattice relaxation time T$_\mathrm{1e}$\cite{Hovav-2013}.

\begin{figure}
\includegraphics[width = 0.45\textwidth]{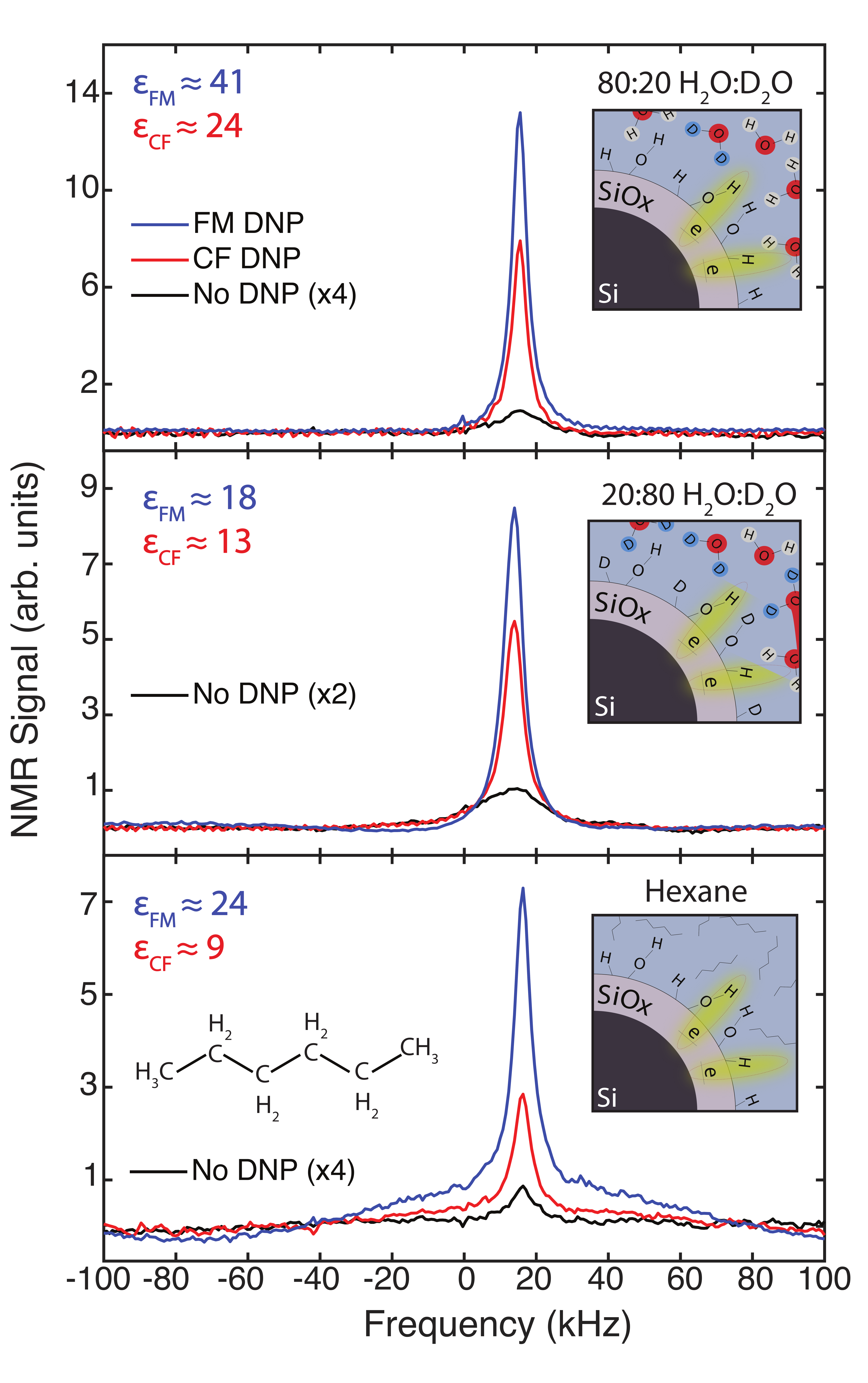}
\caption{Proton NMR spectrum measured without DNP; with CF DNP; and with FM DNP for silicon microparticles suspended in (top) 80\% H$_2$O and 20\% D$_2$O (H$_2$O-rich), (middle) 20\% H$_2$O and 80\% D$_2$O (D$_2$O-rich), (bottom)100\% hexane.  The FM DNP spectra were taken with a modulation amplitude of 30 MHz and a modulation frequency of 10 kHz. The enhancements reported in the figure are for the narrow Lorentzian peak.}
\label{fig:Fig3}
\end{figure}

\subsection{DNP of suspended particles}
We probed the interaction of the silicon powders with three different solvents by first measuring DNP-enhanced proton NMR spectra from the freshly suspended particles, and then repeating the experiments five days later to characterize any interaction-induced changes.

\subsubsection{Freshly-prepared samples}

Figure \ref{fig:Fig3} shows the proton NMR signal measured at 4.2~K with and without DNP when the silicon microparticles are suspended in different solvents: (top) 80\% H$_2$O and 20\% D$_2$O (deuterated or heavy water); (middle) 20\% H$_2$O and 80\% D$_2$O; and (bottom) 100\% hexane.  The ratio of silicon to solvent was 3:1 by weight in all cases.  These experiments were run within a few hours of the samples being prepared.   All of the samples showed a similar narrow (5.1--5.5~kHz) Lorentzian resonance. 
The very wide resonance line expected from the frozen H$_2$O-rich sample was not observed due to the relatively long dead-time (15.5 $\mu$s) of the spectrometer (see Methods).  The D$_2$O-rich sample and the hexane sample did show an additional broader Gaussian-shaped resonance from the frozen solvent.  The width of the Gaussian peak from the solvent protons was 13.1~kHz for the D$_2$O-rich sample and 104~kHz for the hexane sample.  

Deuteration of the solvent reduces the strength of the average proton dipolar couplings as deuterium is a spin-1 nuclear isotope whose gyromagnetic ratio is approximately one-tenth that of the protons.  The solvent and surface peaks partially overlap in the D$_2$O-rich sample but can clearly be distinguished (middle figure).  
Fitting the spectra to the sum of two components allowed us to report DNP enhancements for each component.  The enhancements reported in Figure \ref{fig:Fig3} correspond to those obtained for the narrow Lorentzian peak.  These enhancements are lower than those observed for the dry powders, potentially due to the presence of competing polarization transfer pathways in the solvent, or changing millimeter wave amplitudes due to changes in the dielectric sample loading of the millimeter waves.   The enhancements reported here were obtained at a fixed millimeter wave irradiation time of 256 s, and do not correspond to the maximum steady-state DNP signal.

\begin{figure}
\includegraphics[width = 0.48\textwidth]{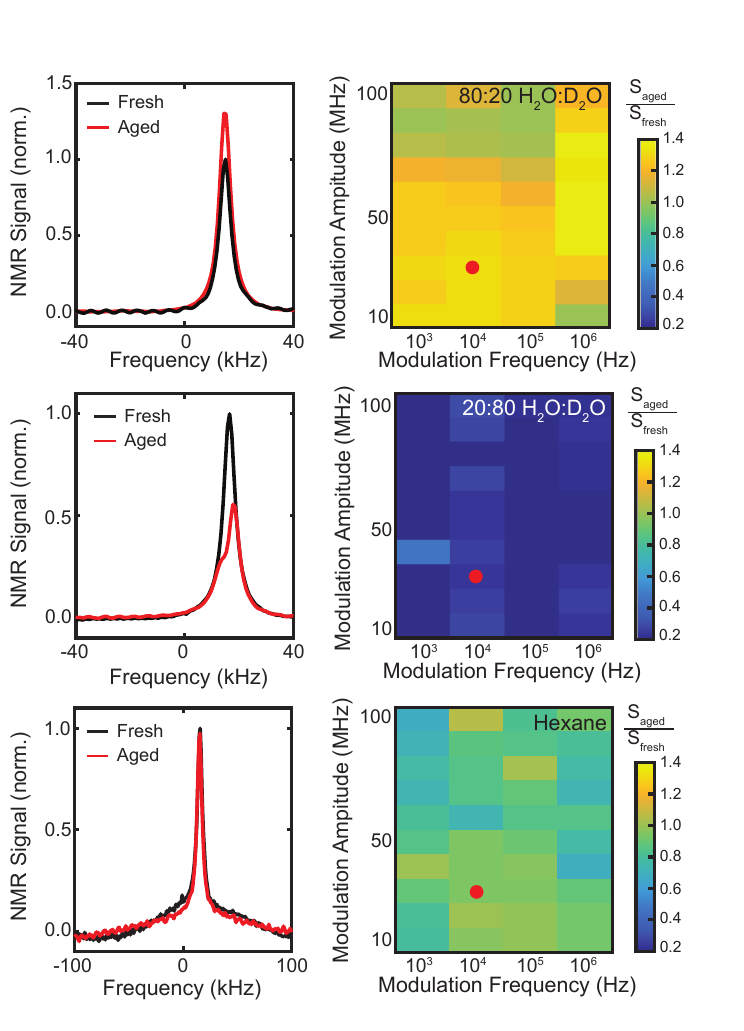}
\caption{Comparison of the proton NMR spectrum obtained with FM DNP for the freshly prepared and the aged samples: (top)  80\% H$_2$O and 20 \% D$_2$O; (middle) 20\% H$_2$O  and 80\% D$_2$O; (bottom) hexane.  Note that the frequency-axis of the H$_2$O-rich and D$_2$O-rich spectra has been expanded to enable better visualization of the spectra.  The FM DNP spectra were taken with a modulation amplitude of 30 MHz and a modulation frequency of 10 kHz. The image plots on the right show the ratio of the amplitude of the narrow peak in the aged sample to the amplitude of the narrow peak in the freshly-prepared sample for all the frequency modulation parameters.}
\label{fig:Fig4}
\end{figure}

The solvent peak enhancements for the D$_2$O-rich sample were $\epsilon_\mathrm{CF}=1.3$ and $\epsilon_\mathrm{FM}=4.2$, and $\epsilon_\mathrm{CF}=5.8$ and $\epsilon_\mathrm{FM}=16.5$ for the hexane sample.  These solvent peak enhancements are reduced compared to the narrow signal since only a small fraction of these molecules are in hyperfine contact with the electron spins \cite{Cassidy-2013a}, and spin diffusion is required to transfer the polarization from the vicinity of the surface to the bulk \cite{Ramanathan-2008}.  Spin diffusion is expected to be faster for hexane given the stronger dipolar coupling.  Additionally, hexane is likely to form glassy matrix on freezing which prevent aggregation of the particles along crystallite boundaries, leading to larger DNP enhancements \cite{Barnes-2008}.  Note that care needs to be taken when interpreting the enhancements for the broad solvent signals as they are very sensitive to slight changes in the applied first-order phase correction, especially for the very broad hexane resonance.

\subsubsection{Aged samples}

Following each of the above experiments the samples were warmed up from 4 K to room temperature over the course of several hours and then stored under ambient conditions for 5 days before being placed back in the cryostat and cooled down for follow-up experiments.  Figure \ref{fig:Fig4} compares the FM DNP enhancement for the freshly prepared samples and the aged samples in each case. 

For the H$_2$O-rich sample (top), the intensity of the narrow peak is seen to increase by a factor of 1.4, indicating significant growth in the bound proton component.   The image plot on the right indicates that this increase was observed with almost all modulation parameters used.  Changes in local chemistry can also change the features of the overall modulation dependence of the signal as discussed earlier.  No change was observed in the linewidth of the signal, indicating that the protons remain well-isolated from each other, even as the overall number of surface protons has increased.

For the D$_2$O-rich sample (middle), the intensity of the narrow peak was seen to drop significantly, and appeared to both shift in frequency and narrow in width.  We attribute these changes to differential rates of exchange for the two surface-bound species with the deuterium atoms in the solvent.  As the surface protons exchange with deuterium, their relative concentration drops, lowering both the intensity and the linewidth of the spectrum.  This decrease was, again, observed across all the modulation parameters.  Following this exchange between the solvent and the surface the relative proton concentration of the solvent should be higher in the aged samples, and the the frozen solvent signal is indeed observed to increase by a factor of 2 across all modulation parameters (see Supplementary Information).  
Overall, the interaction of D$_2$O with the surface is expected to be very similar to that of H$_2$O, with chemisorption resulting in Si-D and Si-OD bonds \cite{Nishijima-1986}.  Here we can see that the chemisorption is a dynamic process with continual exchange between the surface and solvent species.  The hexane sample (bottom) shows no change in DNP behavior for either the narrow or broad components indicating little interaction between the silicon particles and the solvent.

\section{Conclusions}
We have demonstrated that DNP-enhanced proton NMR can be used to follow changes to the bound hydrogen at the surface of silicon microparticles, following chemisorption of the water molecules on the silicon surface.  The narrow proton peak observed indicates the presence of randomly distributed hydrogen atoms that are relatively well-isolated from each other.  These surface bound protons were clearly distinguishable from the broad solvent resonance even at a deuteration level of 80\%.

To the best of our knowledge, the proton NMR spectrum from bound protons at the surface of silicon micro-particles has not previously been measured.   In previous experiments performed at 2.35 T, we were able to measure DNP-enhanced proton NMR from frozen solvent molecules near the surface of the silicon microparticles, but were unable to see a proton signal from the dry particles \cite{Cassidy-2013a}.  The current experiments, performed at a higher magnetic field of 3.34 T and with higher millimeter wave power, used samples from the same batch of polycrystalline silicon powder as the earlier experiments.  It is possible that chemisorption of water during storage may have increased the abundance of surface protons between the two sets of experiments. 

These experiments demonstrate the promise of DNP-enhanced NMR for the study of chemistry in low-dimensional systems.  The non-destructive measurement technique presented here permits serial studies of the same sample over long times, without the need to either dry the particles or perform any other sample preparations before measurement.  Combining the DNP methods with high-resolution NMR techniques, either multiple-pulse coherent averaging techniques or magic-angle spinning \cite{Haeberlen}, could provide more detailed local information about the ordering of the proton spins at the silicon surface.

\begin{acknowledgement}
This research was supported by the National Science Foundation under CHE-1410504.  We thank Charlie Ciambra of the Dartmouth Chemistry Department for help with the use of the FTIR Spectrometer and Dr.\ Charles Daghlian at the Rippel EM Facility for his help with the SEM imaging.   We also thank Prof.\ Walter Kockenberger and Dr.\ Subhradip Paul at the University of Nottingham, and Dr.\ Daghlian for helpful discussions.
\end{acknowledgement}

\section{Material and Methods}

\subsection{Sample Preparation}
For the DNP NMR experiments with the dry powder we used a 5 mm$^3$ cylindrical sample of the Alfa Aesar silicon particles placed in a glass capillary.

The suspended samples were prepared by mixing 100 mg of silicon powder with either 30 $\mu$l of 20:80 H$_2$O:D$_2$O, 30 $\mu$l of 80:20 H$_2$O:D$_2$O, and 50 $\mu$l of hexane. De-ionized water and a new vial of 99\%-enriched D$_2$O (Cambridge Isotope Labs) were used to prepare the samples. The samples were not de-gassed.  A 1 mm$^3$ volume of the suspension was placed in a glass capillary for the DNP NMR experiments.
The experiments on the fresh samples began immediately after sample preparation.   After the experiment, the samples remained in the cryostat until they reached room temperature ($\approx$ 12 hours), after which they were stored at ambient temperature for 5 days before being measured again.

\subsection{FM waveform generation}
\begin{figure}
\includegraphics[width = 0.45\textwidth]{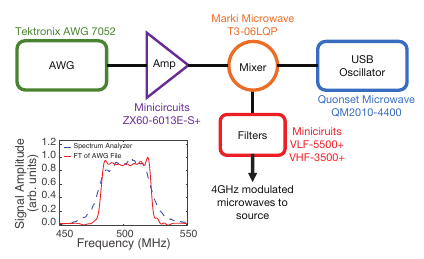}
\caption{Schematic of the frequency mixing stages used to generate the FM waveform.  The plot at the bottom shows a comparison of the fourier transform of the waveform programmed into the AWG, and the frequencies measured on a spectrum analyzer.}
\label{fig:FigS0}
\end{figure}

The 94 GHz FM microwaves are generated using two mixing stages. The modulation is performed in the first mixing stage shown in Figure \ref{fig:FigS0}.  A triangular frequency-modulated waveform is programmed into a Tektronix AWG 7052 with a given $\Delta f$ and $f_\mathrm{m}$. The carrier frequency for the modulated waveform was chosen to be 500 MHz, much less than the AWG's maximum 5Gs/s sample rate to avoid undersampling.

The 500 MHz frequency modulated signal is first amplified (Minicircuits ZX60-6013E-S+ amplifier), and then mixed with a 3.5 GHz carrier generated with a USB-based microwave synthesizer (Quonset Microwave QM2010-4400) using a T3 Marki Microwave mixer (Marki Microwave T3-06LQP).  The output of the mixer is 
 filtered using two Minicircuits filters (VLF-5500+ and VHF-3500+) to eliminate the lower sideband, yielding a triangular modulated signal at 4 GHz. The inset of Figure \ref{fig:Fig5} shows a spectrum analyzer trace of the 500 MHz demodulated signal (blue trace) compared to the Fourier Transform of the file used to program the AWG (red trace). The waveform shown had a modulation bandwidth of 40 MHz and a modulation frequency of 100 kHz. This 4 GHz modulated signal then enters the second mixing stage, where it is mixed with a 90 GHz carrier and filtered to generate a 94 GHz modulated microwave signal which is fed into the DNP probe \cite{Guy-2015}.

\subsection{SEM and FTIR measurements}
The SEM image in Figure~1(b) was taken with a FEI Company XL-30 ESEM-FEG. The silicon particles were mounted to a specimen stage with double-sided tape and loaded into the SEM chamber. The image was acquired with an accelerating voltage of 15 kV and a magnification of 3200x.

The FTIR spectrum in Figure~1(c) was acquired using a JASCO FTIR-6200. A background signal was taken of the chamber without the sample loaded and was used as a background subtraction to the sample spectrum. The sample was then loaded into the chamber after which the chamber was purged with nitrogen gas for 2 minutes. 
Both spectra were acquired using 64 scans at 0.5 cm$^{-1}$ resolution. 
The sample spectrum was baseline corrected by subtracting a 3rd order polynomial fit to the baseline.

\subsection{Data Analysis} 
The NMR spectra in all the figures were generated using matNMR \cite{vanBeek-2007}. For all the spectra, a DC offset correction was first performed using the average of last 51 points (205-256) of the time-domian FID to estimate the baseline.  A left shift of 4 data points was then used to eliminate switching transients and pulse breakthrough effects.  Given the 2.5~$\mu$s dwell time, this contributed an additional 10~$\mu$s to the original 5.5~$\mu$s dead time for the detection.  Following a 1 kHz exponential line broadening, the data were fourier transformed and phased. To determine the appropriate phasing for the spectra, the highest SNR spectrum in an experimental series was phased and the others were phased identically. 

In Figure 1, the thermal spectrum was first attenuated by a factor of 16 to account for differences in experimental receiver gain. 

In Figure 2, the best fit was obtained using a single Lorentzian with a linewidth of 6.2 kHz.  The enhancements shown in the parameter space image of Figure 2 were obtained using the ratio of amplitudes of the corresponding Lorentzian spectral fits.  A high SNR spectrum ($\Delta f = 30$~MHz, $f_\mathrm{m} = 10$~kHz) was first fit to characterize the center and width of the spectra.  These values were then fixed when fitting the amplitudes of all the other FM DNP spectra, the CF DNP spectrum and the thermal signal.  The parameter space explored for the FM DNP contains a total of 50 data points, with the modulation amplitude ranging from 10--100 MHz in increments of 10 MHz and the modulation frequency ranging from 1 kHz to 10 MHz in powers of 10.  The LOD-ESR spectra were smoothed using the Matlab smooth function (moving average filter with a span of 100) to reduce the noise level. 

In Figure 3, the FM DNP spectra for the H$_2$O rich sample was fit to a single Lorentzian (the very broad component is unseen because of the left shift of 4 data points). The width of the Lorentzian was 5.2 kHz, slightly narrower than that observed for the dry powder.  The D$_2$O rich sample and the hexane samples were fit using the sum of a narrow Lorentzian and a broad Gaussian. For the D$_2$O rich sample, the broad Gaussian had a linewidth of about 13.1 kHz and the narrow Lorentzian had a linewidth of about 5.1 kHz. For the hexane sample, the narrow Lorentzian had a linewidth of about 5.4 kHz and the broad Gaussian had a linewidth of 104 kHz. 
Once again, a high SNR spectrum ($\Delta f = 30$~MHz, $f_\mathrm{m} = 10$~kHz) was first fit to characterize the center and width of each of the spectral peaks.  These values were then fixed when fitting the amplitudes of the other spectra.  The magnitudes of the fits for each component were then used to calculate the reported DNP enhancements. The Supplementary Information provides additional details on the spectral fits.

In Figure 4, the amplitudes of the aged spectra were fit using the same center and width parameters determined from the freshly prepared samples.  Excellent fits were obtained in all cases --- see Supplementary Information.
The ratios of the fits to the narrow spectral line were then used to create the maps shown in the figure.
Note that we observed small frequency shifts in the spectral peaks in different experiments, and occasionally observed the peaks to shift during a single cool down.  We determined that changes in sample placement within the coil, and rotational variations in the probe position could lead to changes of line position by 1 kHz and 600 Hz respectively, due to the inhomogeneity of the magnetic field.  The peak shifts during a cool down were due to changes in helium pressure shifting the sample within the coil.  In order to compare the spectra from different experimental cooldowns, we shifted the centers of the aged peaks in Figure 4 to ensure overlap.  The H$_2$O-rich spectrum was shifted by 500 Hz, the D$_2$O-rich spectrum was shifted by 1 kHz, and the hexane spectrum was shifted by 500 Hz. 

\section{Supplementary Data}

\subsection{Dry baseline signal}
\begin{figure}
\includegraphics[width = 0.75\textwidth]{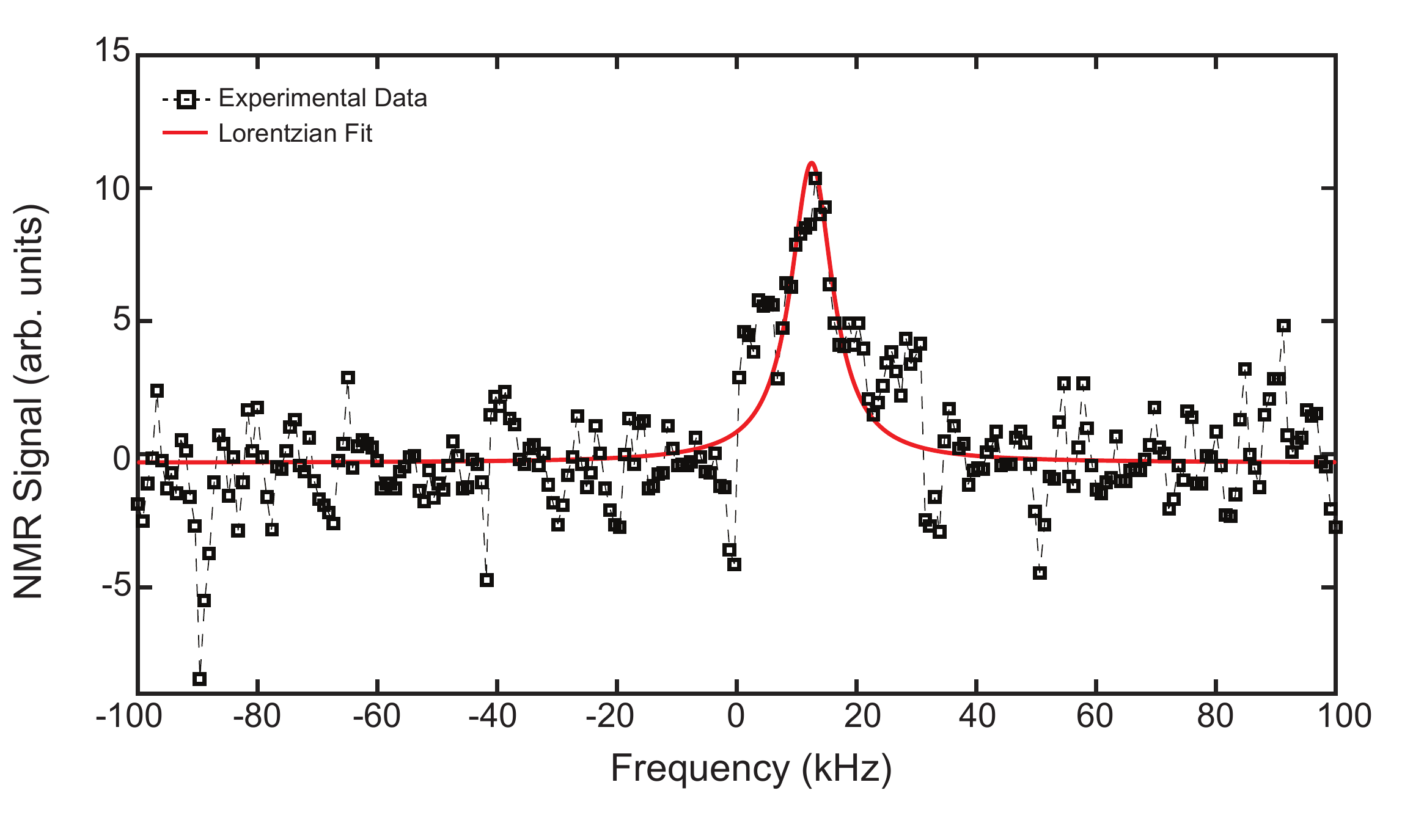}
\caption{Baseline signal for the dry silicon powder. The data (black squares) is fit to single Lorentzian resonance (red trace) with a linewidth of 6.2 kHz.}
\label{fig:FigS1}
\end{figure}
Figure \ref{fig:FigS1} shows the single Lorentzian fit to the NMR spectrum of the dry powder in the absence of applied microwaves (no DNP).   A high SNR DNP-enhanced proton NMR spectrum ($\Delta f = 30$~MHz, $f_\mathrm{m} = 10$~kHz) was first fit to characterize the center and width of the spectra (shown in Figure 2 of the main paper).  These values were then fixed when determining the best fit to the baseline spectrum -- shown in red in the figure.  The signal to noise for the baseline spectrum is seen to be about 4, resulting in some uncertainty in the reported DNP enhancements.

\subsection{Two-component fits}
\begin{figure}
\includegraphics[width = 0.55\textwidth]{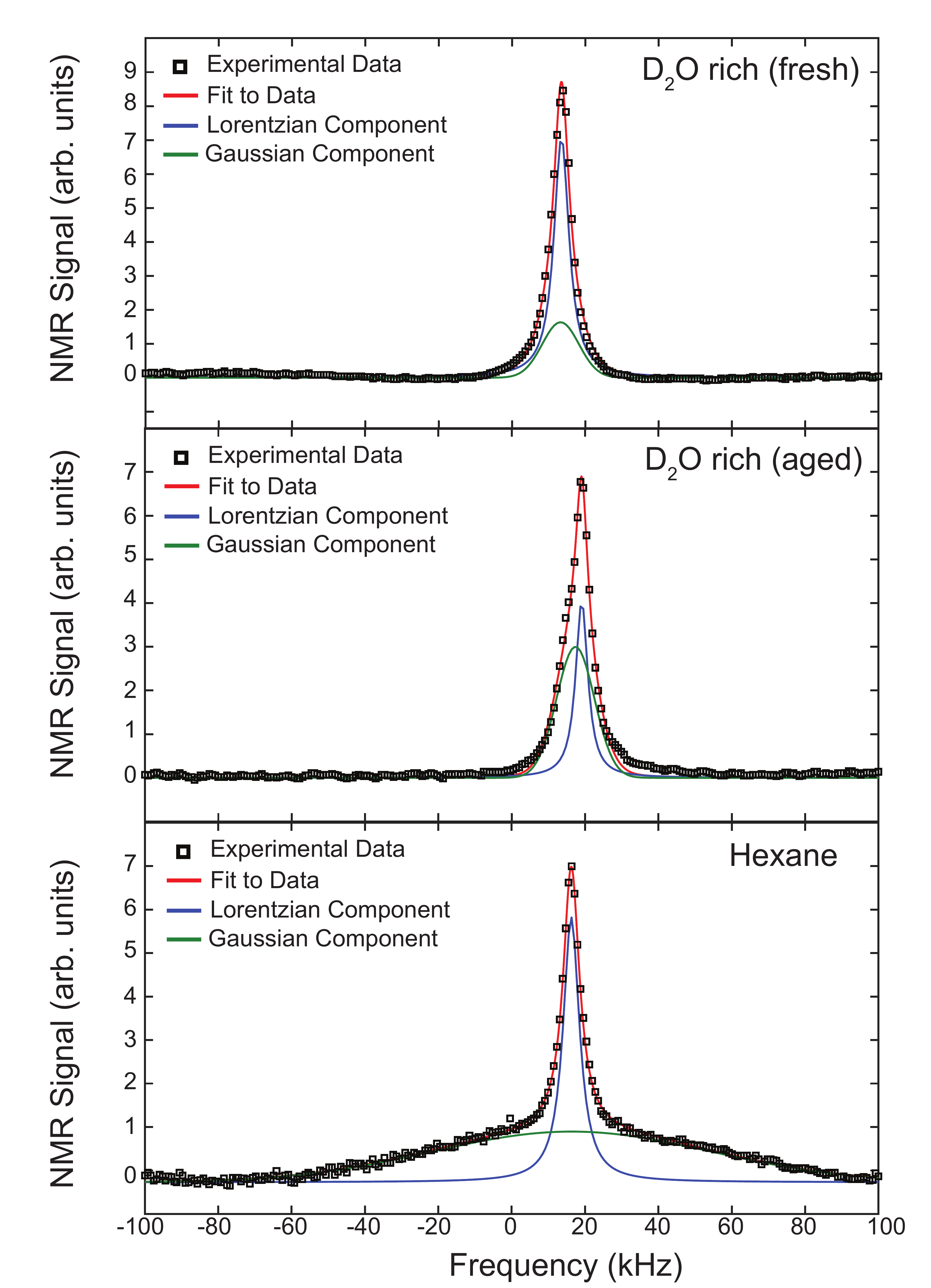}
\caption{Proton NMR spectra and two-component spectral fits to the broad and narrow resonances for suspended silicon microparticles: (top) D$_2$O-rich solvent - freshly prepared;  (middle) D$_2$O-rich solvent - aged; (bottom) hexane solvent. The plots show the experimental data (black squares) and total fit (red trace), along with the decomposed broad (green trace) and narrow (blue trace trace) components of the fit.  These spectra were obtained using FM DNP with $\Delta f = 30$~MHz and $f_\mathrm{m} = 10$~kHz.}
\label{fig:FigS2}
\end{figure}

Figure \ref{fig:FigS2} shows the two-component fits for to NMR spectra for the fresh and aged D$_2$O-rich sample and the hexane sample obtained with FM DNP.  The data were fit with two components: a narrow Lorentzian line and a broad Gaussian line. For each sample, the linewidth of the narrow component was observed to match that of the dry powder ($\sim$ 5--6 kHz) while the broad component was observed to change with the solvent. For the D$_2$O-rich sample, the broad component was about 13.1 kHz and the narrow component was about 5.1 kHz for the fresh sample.  The width of the narrow component decreased in the aged samples (4.2 kHz), but the width of the broad component was relatively unchanged. For the hexane sample, the linewidth of the broad Gaussian had a linewidth of 104 kHz.

\subsection{Growth of the DNP signal}
\begin{figure}
\includegraphics[width = 0.75\textwidth]{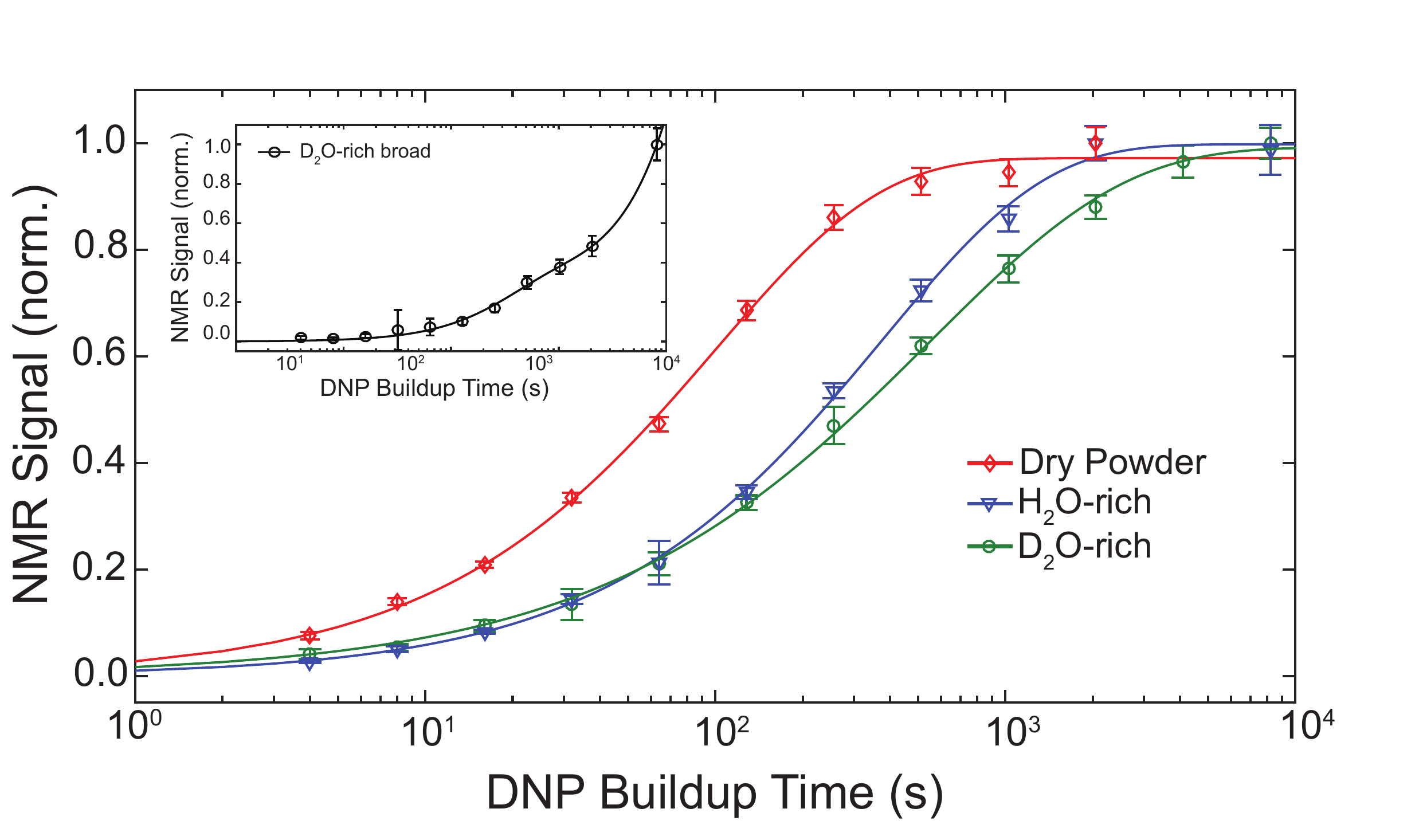}
\caption{Buildup curves for dry powder (experimental data open diamonds, stretched fit red line) and narrow components of H$_2$O-rich (experimental data open triangles; stretched fit blue line) and D$_2$O-rich (experimental data open circles, stretched fit green line) samples. The inset shows the growth of the broad component of the D$_2$O-rich sample, fit with a double exponential.}
\label{fig:FigS3}
\end{figure}

We also measured the growth of the DNP signal as a function of millimeter wave irradiation time. Figure \ref{fig:FigS3} shows the growth for the dry powder and for the narrow components for the freshly prepared samples of powder dispersed in mixtures of water and D$_2$O.  For each of these, the best fit was obtained with a stretched exponential, similar to those obtained in recent $^{13}$C DNP experiments with diamond nanoparticles \cite{Rej-2015}. For the dry powder, the stretching exponent was $0.77  \pm .07$ and the characteristic time constant was measured to be T$= 35\pm 10$ s. For the H$_2$O-rich and D$_2$O-rich samples the stretching exponents were  $0.64 \pm .03$  and $0.77 \pm .05$, respectively, and the characteristic time constants were T$= 58\pm 10$ s for the  H$_2$O-rich and T$= 98\pm 30$ s for the D$_2$O-rich sample, significantly longer than the dry powder.  The stretched exponential growth of the narrow component is likely due to a distribution of electron-nuclear interaction strengths, as spin diffusion effects are unlikely to play a role in DNP here.

The inset of Figure \ref{fig:FigS3} shows the buildup curves for the broad component of the freshly prepared D$_2$O-rich sample. This was best fit with a sum of two exponentials. The time constants were $204.9\pm 92.4$ s ($\sim$22 \%) and $4168 \pm 1599$ s ($\sim$78 \%) for the D$_2$O-rich sample.  The bi-exponential behavior is due to different compartments of spins at different distances to the dangling bond defects. Solvent spins that are closer to the surface of the particles are in direct hyperfine contact with the dangling bond electron spins and experience faster polarization buildup, which those located farther from the particle surface, rely on spin diffusion to polarize.

\begin{figure}
\includegraphics[width = 0.55\textwidth]{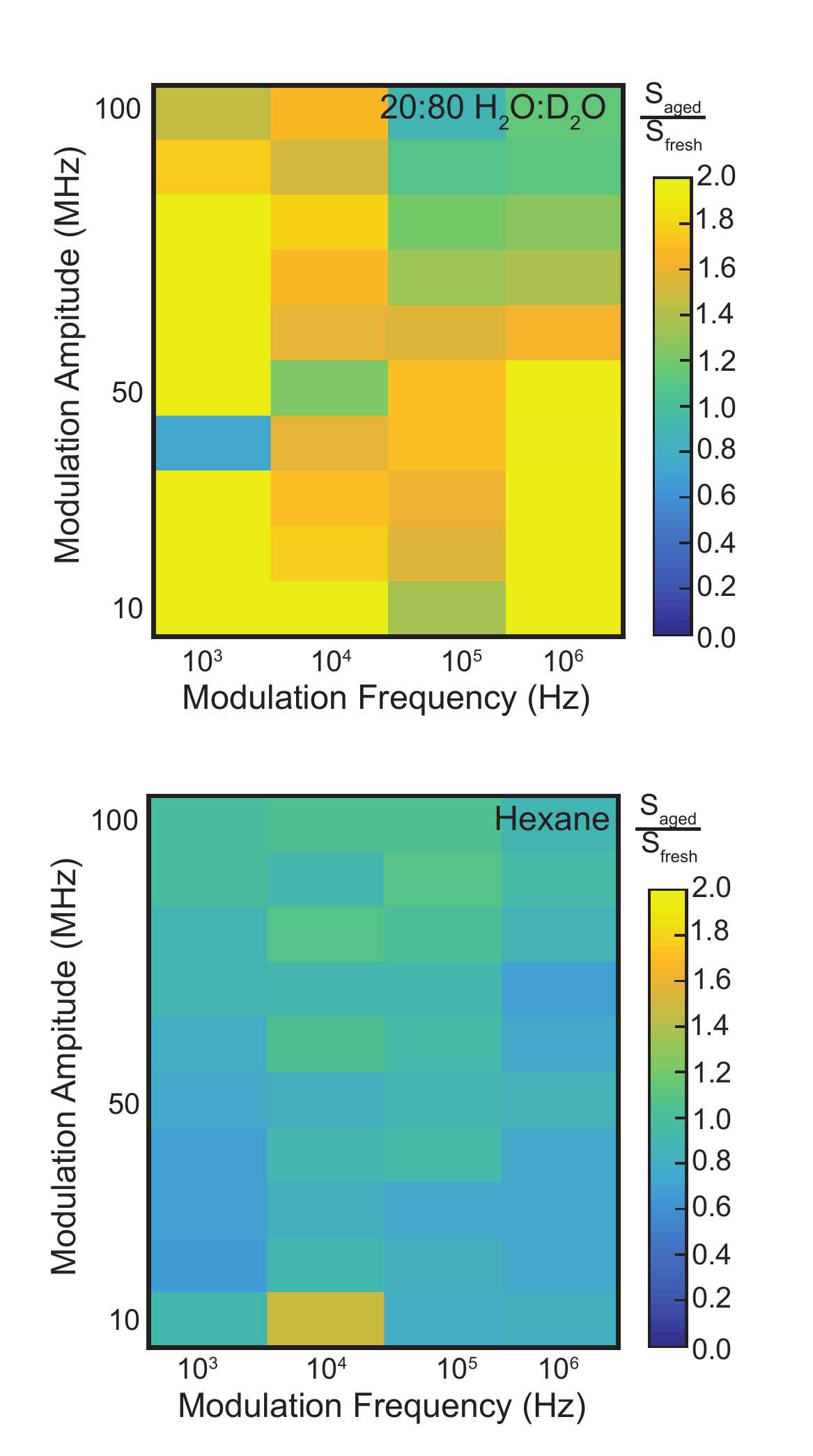}
\caption{Image plot showing the (top) the change in the broad component for the D$_2$O-rich sample and (bottom) for the hexane sample. The broad component for the D$_2$O-rich sample increases by about a factor of 2 while the broad component for the hexane sample remains unchanged.}
\label{fig:FigS4}
\end{figure}

\subsection{Broad Component Behavior}

Figure  \ref{fig:FigS4} shows image plots of the ratio of the amplitude of the broad peak in the aged sample to the amplitude of the broad peak in the freshly-prepared sample for all the frequency modulation parameters, for both the D$_2$O-rich sample (top) and the hexane sample (bottom). The broad component for the  D$_2$O-rich sample sample is seen to increase by about a factor of two during aging, as the surface protons enter solution following exchange with the deuterium.  The broad component of the hexane sample remains unchanged, consistent with the narrow component also remaining unchanged. The ratios were calculated using the amplitudes of the fits to the broad spectral components of both the fresh and aged spectra.

\subsection{Signal-to-Noise Ratios}
\begin{figure}
\includegraphics[width = 0.45\textwidth]{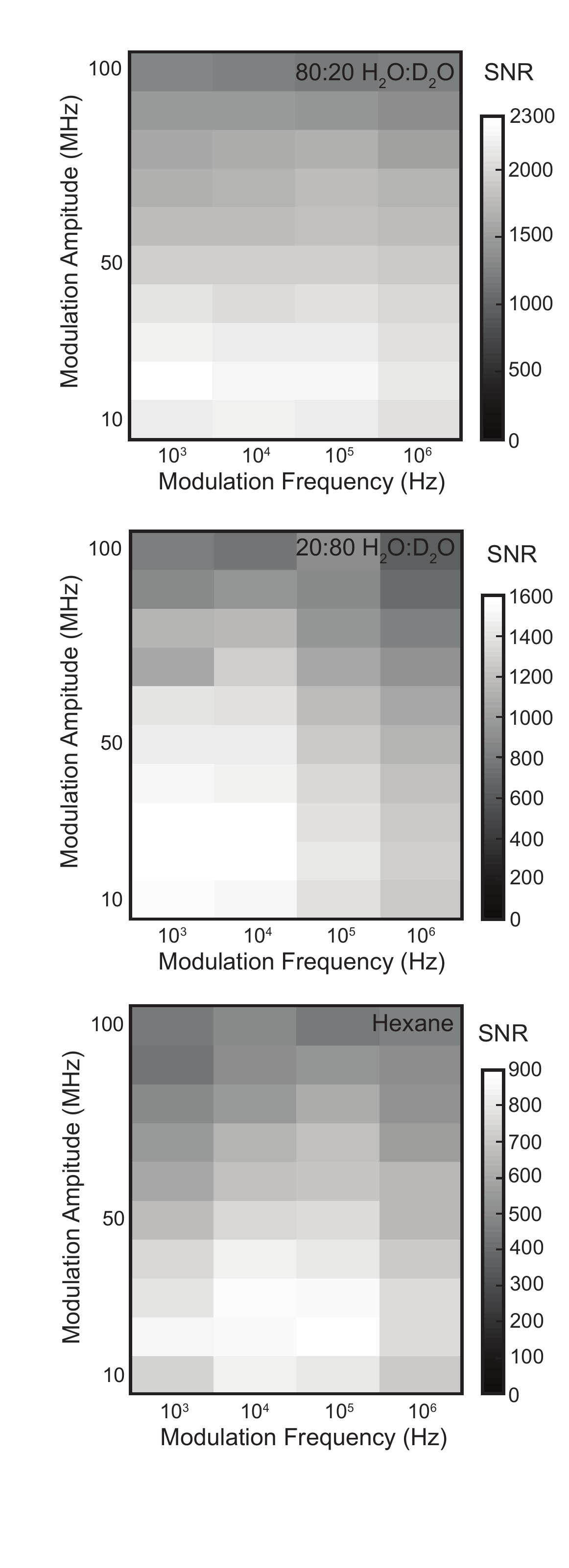}
\caption{Image plot of the signal-to-noise ratio for proton NMR signal: (top) H$_2$O-rich sample; (middle)  D$_2$O-rich sample; and (bottom) hexane sample.}
\label{fig:FigS5}
\end{figure}

Figure \ref{fig:FigS5} shows the signal-to-noise ratios for each of the freshly-suspended samples.  The standard deviation of the first 10 points in the spectral domain of the baseline signal was used to characterize the noise.  This standard deviation was observed to be remain the same across almost all the spectra measured.  
In some measurements we observed random transient spikes in the time domain - which led to oscillatory features in the baseline, preventing accurate estimates of the noise in individual experiments.  The enhancement pattern is similar to that observed with the dry powder (see Figure 2(b) of the main paper).

\bibliography{SurfaceDNP}

\end{document}